%% file: discrimination.tex
\documentclass[manuscript,screen]{acmart}
\AtBeginDocument{%
  \providecommand\BibTeX{{%
    \normalfont B\kern-0.5em{\scshape i\kern-0.25em b}\kern-0.8em\TeX}}}

\setcopyright{acmcopyright}
\copyrightyear{2020}
\acmYear{2020}
\acmDOI{https://doi.org/10.1145/3313831.XXXXXXX}

\acmJournal{JACM}
\acmVolume{37}
\acmNumber{4}
\acmArticle{0} % TO-DO maybe change this at the time of submission
\acmMonth{8}

\usepackage{csquotes}
\usepackage{multirow}
\usepackage{graphicx}
\usepackage{colortbl}
\usepackage{arydshln}
\usepackage{wrapfig}
\usepackage{sidecap} % side caption for figures
\usepackage{xcolor} % text coloring

\usepackage{microtype}        % Improved Tracking and Kerning
\usepackage{ccicons}          % Cite your images correctly!
\usepackage{todonotes}
\usepackage[shortlabels]{enumitem}

\def\plainkeywords{Discrimination; Coping; Mental Health; Education; Social Support; Intervention; Interview; Contextual Inquiry; Experience Sampling,}

\newcommand{\fig}[1]{Figure~\ref{#1}}
\newcommand{\tbl}[1]{Table~\ref{#1}}

\newcommand{\sect}[1]{Section~\ref{#1}}

\newcommand{\eg}[0]{{e.g.,~}}
\newcommand{\ie}[0]{{i.e.,~}}
\newcommand{\etc}[0]{{etc.}}
\newcommand{\qt}[1]{\textit{``#1''}}

\newcommand\numuwii[0]{253 }

\newcommand\numdiscrimination[0]{90 }
 
\newcommand\numnotinterviewee[0]{76 } % 90 reported everyday discrimination; 14 were interviewed
\newcommand\numinterviewees[0]{14 }

\newcommand\numintervieweesnresolved[0]{9 }

\newcommand\TrajecoryContinued[0]{P480}

\newcommand\DirectClass[0]{P326, P471, P480, P488}
\newcommand\DirectProgram[0]{P362}
\newcommand\DirectMajor[0]{P307, P411} 
\newcommand\IndirectGender[0]{P404, P521, P527, P552}

\newcommand\ResolvedDirectNotConfidence[0]{P326}
\newcommand\ResolvedIndirectPosTONeg[0]{P521}
\newcommand\ResolvedDirectPosTONeg[0]{P488}
\newcommand\NResolvedDirectNegTOPos[0]{P307} 
\newcommand\NResolvedIndirectDelay[0]{P404}
\newcommand\ResolvedDirectDelay[0]{P471}
\newcommand\ResolvedDirect[0]{P471}

\newcommand\ResolvedIndirectInsignificant[0]{P420}

\newcommand\LowCapacityIntellect[0]{P480}

\newcommand\HighCapacity[0]{P471}
\newcommand\LowOutcomeEfficacy[0]{P480}
\newcommand\NeutralOutcomeEfficacy[0]{P552}

\newcommand\SocialConcernEGii[0]{P527}

\newcommand\NoSocialConcern[0]{P521}

\newcommand\SilentCode[0]{P480, P521, P527}

\newcommand\SilentCodeiii[0]{P527}

\newcommand\NResolvedUpwardHelpii[0]{P362}

\newcommand\ResolvedUpwardHelpi[0]{P471}

\newcommand\ResolvedCommiserate[0]{P521}
\newcommand\NResolvedCommiserate[0]{P307, P404, P411, P480, P527, P552}

\newcommand\NResolvedCommiserateEGii[0]{P411}

\newcommand\NResolvedADistraction[0]{P307}
\newcommand\ResolvedADistraction[0]{P326}

\newcommand\AvoidanceNegii[0]{P404}

\newcommand\AvoidancePosi[0]{P411}

\newif\ifoutline
\outlinefalse
\definecolor{darkblue}{RGB}{0,0,139}

\newif\ifsensor
\sensorfalse  % Uncomment if not reporting on sensors, comment otherwise

\newif\ifcommentsvisible
\commentsvisiblefalse  % Uncomment if you do *not* want to see the comments, comment otherwise

\makeatletter
\def\contentsname{Contents}
\def\tableofcontents{%
    \section*{\MakeUppercase{\contentsname}}%
    \@starttoc{toc}%
    }
\makeatother

\definecolor{ysspurple}{RGB}{102,0,102}
\definecolor{yssdgreen}{RGB}{0,100,0}
\definecolor{yssteal}{RGB}{0,128,128}
\definecolor{yssnavy}{RGB}{0,0,128}
\definecolor{ysspink}{RGB}{255,20,147}
\definecolor{yssgreen}{RGB}{107,142,35}
\definecolor{yssgray}{RGB}{47,79,79}
\definecolor{yssbluepurple}{RGB}{138,43,230}

\newcommand\discrimination{perceived discrimination}
\newcommand\Discrimination{Perceived discrimination}

\newcommand\rev[2]{\label{#1}\textcolor{black}{#2}}
\begin{document}

\title[Examining Needs and Opportunities for Supporting Students Who Experience Discrimination]{Examining Needs and Opportunities for Supporting Students Who Experience Discrimination}

%\author{Anonymous Author(s)}
\author{Yasaman S. Sefidgar}
\orcid{0000-0001-8990-699X}
\email{einsian@cs.washington.edu}
% \affiliation{%

\author{Paula S. Nurius}
% \affiliation{%
\email{nurius@uw.edu}

\author{Amanda Baughan}
% \affiliation{%
\email{baughan@cs.washington.edu}

\author{Lisa A. Elkin}
% \affiliation{%
\email{lelkin@cs.washington.edu}

\author{Anind K. Dey}
% \affiliation{%
\email{anind@uw.edu}

\author{Eve Riskin}
% \affiliation{%
\email{riskin@ee.washington.edu}

\author{Jennifer Mankoff}
\email{jmankoff@cs.washington.edu}

\author{Margaret E. Morris}
\email{margiemm@uw.edu}
\affiliation{%
  \institution{University of Washington}
%  \country{USA}
}

%\renewcommand{\shortauthors}{Anonymous Author(s)}
\renewcommand{\shortauthors}{Yasaman S. Sefidgar et al.}

\begin{abstract}
\rev{}{\Discrimination} is common and consequential. \rev{}{Yet, little support is available to ease handling of these experiences. Addressing this gap, we report on a need-finding study to guide us in identifying relevant technologies and their requirements. Specifically, we examined} unfolding experiences of \rev{}{\discrimination\ among college students and found factors to address in providing meaningful support. W}e used semi-structured retrospective interviews with \numinterviewees students to understand their perceptions, emotions, and coping in response to discriminatory behaviors within the prior ten-week period. These \numinterviewees students were among \numdiscrimination who provided experience sampling reports of unfair treatment over the same ten-week period. We found that discrimination is more distressing if students face related academic and social struggles or \rev{}{when the incident triggers} beliefs of inefficacy. \rev{}{We additionally identified patterns of effective coping. By grounding} the findings \rev{}{in} an extended stress processing framework, we \rev{}{offer a principled approach to intervention design, which we illustrate through} incident-specific and proactive intervention paradigms. 
\end{abstract}




\maketitle

\input{1-intro}
\input{2-back}

\input{3-method}
\input{4-analysis}
\input{5-result}

\input{6-quant}
\input{7-discussion}

\input{8-conclusion}
\bibliographystyle{ACM-Reference-Format}
\bibliography{discrimination}

% %%
% %% If your work has an appendix, this is the place to put it.
% \appendix









\input{9-appendix}

\end{document}

%% file: 1-intro.tex
\section{Introduction}
\label{sec:intro}
\rev{}{We report on a need-finding study to guide the design of technologies that can support individuals who perceive discriminatory behaviors in their social life within an academic context. 
\Discrimination\ experiences are important to address because they are common and consequential. One in four adults in the U.S. \cite{aaa-stress} report these experiences, which} trigger affective and physiological responses that increase the risk for health complications \cite{Smart:2010}. 
In addition to gradually manifesting negative physical and mental health outcomes \cite{Kessler:1999, Pavalko:2003}, poor task performance, with measurable organizational ramifications have been linked to \rev{}{\discrimination} \cite{OBrien:2016}. These concerns are particularly pertinent for college students because they report \rev{}{\discrimination\ at higher rates} \cite{Boysen:2009} and college years are developmentally critical; many students transition from adolescence into adulthood during this period. Moreover, 75\% of lifetime mental health concerns of the kind linked to \rev{}{\discrimination} start during this time frame \cite{Kessler:2005}. Understanding \rev{}{design} requirements \rev{}{for delivering} effective interventions in this population is thus critical and high-impact. 

Technology interventions to support individuals who \rev{}{\discrimination} has recently received attention. For example, \citet{To:2021} elicited ideas for technologies that \rev{}{can} help individuals cope with \rev{}{inter-personal} racism experiences. Through participatory design workshops, the researchers identified uncertainty reduction and emotional comfort as areas for technology interventions. Overall, however, research that can guide the design of discrimination-specific technologies is still in its infancy. \rev{}{There is little principled guidance on what technologies we should design and the requirements these technologies should satisfy. Our work addresses this gap by contributing knowledge of how discrimination experiences unfold and what factors influence their evolution. As such, it puts forth a systematic approach to intervention design: targeting the influencing factors is a critical first step in working toward designing interventions that deliver meaningful and extended support}.

We present qualitative and quantitative studies that offer a detailed analysis of situational processing of \rev{}{discriminatory} encounters and coping with them over time. More concretely, we interviewed \numinterviewees first and second year university students who \rev{}{\discrimination} in the developmentally critical college setting. Students in the study reported a range of differential treatment experiences for reasons including gender, race, and socioeconomic status, within ten weeks of the interview session. We evaluated interview observations using experience sampling data over the same ten-week period, collected from the interviewees and \numnotinterviewee additional students. Our contributions are: 

\begin{enumerate}
    \item We identify factors that amplify the distress of \discrimination. We find that distress is amplified when incidents relate to an ongoing social or academic challenge. The distress may not resolve until that underlying challenge resolves. In addition, we find incidents are especially deleterious when they activate existing beliefs about not being competent or not having control. 
    \item We \rev{}{highlight} patterns of effective coping, particularly in the form of persistent, targeted, and diverse active coping strategies. We also report on barriers students face in coping including beliefs of social rejection.
    \item We interpret these findings within an extended stress processing framework \rev{}{and offer a principled approach to identifying relevant technologies and their requirements.} Specifically, we consider both \textit{incident-specific} interventions to aid coping with individual incidents and \textit{proactive} interventions in social and educational structures to help resource development independent of particular incidents.
  \end{enumerate}

The following section presents the background on \rev{}{\discrimination} and its negative consequences as well \rev{}{as} the \rev{}{Human-Computer Interaction (HCI)} research related to this topic. We additionally present a theoretical framework relevant to the study of stress and coping in the context of discrimination. We explain our interview and experience sampling methodology and share findings in Sections~\ref{sec:study}-\ref{sec:study-ema}. In \sect{sec:discussion}, we discuss insights from our qualitative and quantitative analysis, theoretically \rev{}{ground} our observations, and use this \rev{}{grounding} to inspire design paradigms and a range of technology solutions. 

%% file: 2-back.tex
\section{Background and Related Work}
\label{sec:back}
We motivate our research by defining \rev{}{\discrimination} and reviewing its consequences, particularly in the context of higher education. We then explore the design of support technologies in relation to experiences of discrimination.
This review highlights the importance of characterizing the unfolding experiences of discrimination.  
We next describe psychological theories of stress that are relevant to the study of unfolding discrimination experiences. 

\subsection{Discrimination and its Impact on Health}
\label{sec:back-microaggression-impact}

\rev{}{The focus of our research is on \discrimination\ experiences. We acknowledge the structural dimensions of discrimination and the systematic oppression of certain social groups. However, the present work primarily attends to and examines experiences that are reported by the targets as differential and unfair. We chose this focus because it is the target's perception, independent of objective verification, that is linked to distress \cite{aaa-stress, Pascoe:2009}, and it is our desire to provide support for this kind of experience. 
Our choice of focus area by no means indicates that discrimination is \textit{just} perceived. The reality of these experiences has been documented over and over (\eg \cite{Moss-Racusin:2012}). Moreover, all discriminatory encounters have their roots in prejudice, stereotypes, and other forms of bias that are typically systematic \cite{Pager:2008}.}

\rev{}{We use \citet{Pascoe:2009}'s definition for \discrimination} as ``a behavioral manifestation of a negative attitude, judgment, or unfair treatment toward members of a group''. Discrimination is often based on race, gender, sexual orientation, or socioeconomic status. However, it is not inherently tied to \rev{}{these} social structures. The concept essentially ``implies a rejection or exclusion of the targeted group'' and communicates the perpetrator's negative beliefs about members of that group \cite{Schmitt:2014}. As such, discrimination has been studied more broadly, for example in relation to mental illness or weight \cite{Schmitt:2014, Potter:2019} and among dominant religious groups~\cite{Hyers:2008}. Past research has shown that ``the generic experience of discrimination generates psychological distress regardless of the attribution and the characteristics of the target'' \cite{Williams:2012} and suggests the same underlying psycho-physiological processes apply independent of the type of discrimination \cite{Smart:2010}. This is not to say that discrimination impact is comparable across social groups; it is undeniable that minoritized groups more frequently experience discrimination with more severe outcomes \cite{Schmitt:2014}. Generic reactions and underlying response processes only suggest that investigating a wide variety of discriminatory experiences may yield insights into how people generally react to \discrimination.

As a stressor, \discrimination\ influences mental and physical health \cite{Pascoe:2009}. Discriminatory events can evoke affective and physiological responses ``that may increase vulnerability to pathogenic processes'' \cite{Smart:2010}. Higher exposure to \discrimination\ is associated with increased risk for depression and anxiety \cite{Kessler:1999} as well as cardiovascular disease \cite{Lockwood:2018}. The more strongly these experiences elicit feelings of powerlessness and lack of control over life outcomes, the more severe are the outcomes \cite{Schmitt:2014}. Moreover, repeated exposure to \discrimination\ events may prompt maladaptive responses, such as withdrawal and substance abuse with additional adverse consequences \cite{Pascoe:2009}.

Much past research is cross-sectional and primarily documents the long-term negative health outcomes associated with \discrimination\ \cite{Kessler:1999, Williams:1997}. These outcomes develop over time \cite{Pavalko:2003} but it is unclear how \cite{Potter:2019}. A detailed understanding of the short-term dynamics of the experiences is needed to further the insights into the association between \discrimination\ and the long-term outcomes \cite{Potter:2019} \rev{}{and to elucidate ways to prevent these outcomes}. 

\subsection{Discrimination in Higher Education} 
\label{sec:back-microaggression-ed}
\Discrimination\ is a familiar experience on college campuses, where about half of students report such incidents in classrooms \cite{Boysen:2009}. Performance, in educational contexts \cite{Billingsley:2019, Datu:2018, Craig:2014} and more broadly \cite{OBrien:2016}, is adversely influenced by  discrimination. \citet{Singletary:2009} and \citet{Walker:2021} provide robust evidence for the significant toll of even subtle interpersonal discrimination on performance through a series of studies where they experimentally manipulate discriminatory behaviors and show reduced subjective and objective measures of performance only in the presence of \discrimination. One explanation is that the inherent ambiguity associated with discriminatory behaviors leads to increased cognitive load and attention deficits that deplete individuals of resources to perform at their normal cognitive capacity \cite{Singletary:2009, Walker:2021}. These findings can partially explain the observed performance gaps, \eg among women who under-perform in math and science despite having high abilities \cite{sue2010microaggressions, bell2002sincere, cadinu2005women, gore2000lesbian}. In addition to cognitive load, discrimination undermines self-esteem \cite{Becerra:2020} and creates hostile and invalidating learning climates that impede enrolment, retention, and graduation \cite{OHara:2012, sue2010microaggressions}. 

Considering these severe consequences, effective interventions in the context of higher education is much needed. Designing such interventions depends on identifying aspects of the student experience that can be influenced to mitigate the harm. Understanding how each discriminatory experience unfolds in student lives and what factors influence it helps us in achieving this goal. %Our work provides such understanding.

\subsection{Technology Interventions for Discrimination}
\label{sec:back-tech}
Addressing the needs of people who encounter discriminatory behaviors is important as these experiences are frequent and consequential. Questions of what these needs are and whether technologies can help address them are underexplored. However, research in this space is growing. For example, \citet{To:2021} investigated needs of individuals coping with incidents of racism by eliciting ideas about technology interventions. Through participatory design workshops they found uncertainty reduction and emotional comfort as areas for intervention. This research calls attention to qualities that are often overlooked in technology design. For example, reducing uncertainty is very challenging using existing social technologies \cite{To:2020}. These technologies \textit{can} address this need if properly designed. However, they were \textit{not designed} to do so. The burden of adapting the poorly designed technologies to the specific needs within the context of discrimination then falls on the targets themselves \cite{To:2020}. 

Our aim is to add to \rev{}{the} line of research on identifying needs that should be addressed for effective handling of \discrimination. By characterizing the unfolding experience of very recent discriminatory encounters we can gain detailed insights about meaning-making and coping along with the situational factors that influence these processes. 
This knowledge would allow us to identify needs that 
technologies can target to provide effective support. In the next subsection we review psychological frameworks that we can draw on for the study of how the experience of \discrimination\ unfolds.

\subsection{Linking Short-term Responses to Discrimination to Long-term Behavior} 
\label{sec:back-microaggression-model}

\begin{figure}[h]
    \centering
    \includegraphics[width=0.8\textwidth]{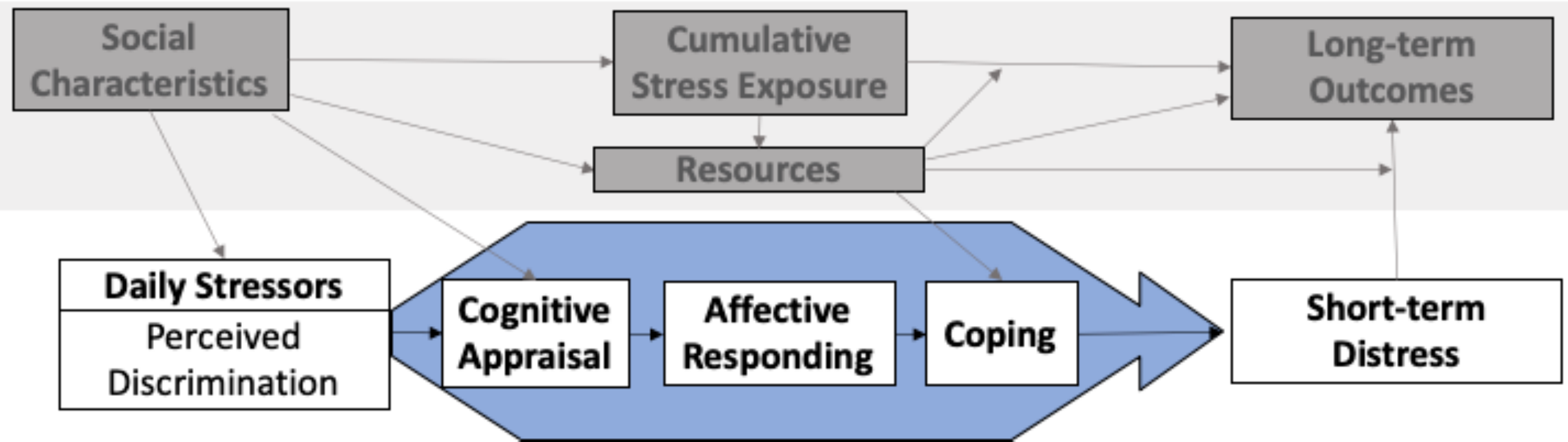}
    \caption[Extended Stress Process Model for Perceived Discrimination]{Extended Stress Process Model for short-term distress and long-term outcomes associated with \discrimination. 
    Individuals engage in CAB processing as they face discrimination. Social characteristics and prior stress exposure influence cognitive-affective processing. Moreover, resources can compound or attenuate behavior response, generating higher or lower levels of stress `load' or `overload.'
    }
    \label{fig:microaggression-model-stress}
\end{figure}

\Discrimination\ has been conceptualized as a stressor (\eg \cite{Ong:2009,Sefidgar:2019}) and as such its relation to health \rev{}{and performance} outcomes has been studied within the stress processing model \cite{Turner:2013} shown in \fig{fig:microaggression-model-stress}, light gray box. Specifically, social characteristics (\eg race, gender, or socioeconomic status) are associated with greater chronic and proliferating stress exposure. In other words, certain groups more frequently experience discrimination and the higher or more chronic the exposure, the more severe the consequences. This model also accounts for the role that resources, personal (\eg coping skills) or social (\eg support networks), play in buffering the impact o\rev{}{f discrimination} (\cite{Pascoe:2009, Schmitt:2014}). 

Many applications of the stress process model in capturing contexts of \discrimination\ have focused on cumulative and, in some cases, life course experiences, explaining \textit{long-term} outcomes. Examinations that account for the short-term impact of \discrimination\ have been undertaken but do not identify the mechanistic elements of the short-term response \cite{Ong:2009}; they primarily demonstrate the connection between the daily incidents and short-term distress, \ie white boxes titled ``Daily Stressors'' and ``Short-term Distress'' in \fig{fig:microaggression-model-stress}. 
Adding to this literature, \citet{Sefidgar:2019} 
showed that discrimination-related stress affects behaviors such as sleep, phone use, number of calls, physical activity, and step count on the day of the event. 
While indicative of potential impact mechanisms, this work does not examine specifics of the connection between discriminatory incidents and short-term distress response.

\rev{}{We posit that t}he framework of {\em C}ognitive, {\em A}ffective, and {\em B}ehavioral (CAB) responding \cite{folkman2011oxford, nurius2013stress} offers a promising approach to understanding distress in response to perceived discrimination, as a stressor, in the short-term (\fig{fig:microaggression-model-stress}, boxes within blue arrow). CAB is a general framework to model cognitive-affective processes activated by stressors: \textit{Cognitive response} represents `meaning making,' \ie subjective perceptions or appraisals of (1) the extent to which a circumstance is benign, poses benefit, or potential harm, and (2) the extent to which the person perceives the threat to be controllable or that they have resources to manage the circumstances. As stressors are processed cognitively, \textit{affective responses}, 
These situational appraisals and subsequent affective states then shape a person's \textit{behavioral responses}. Varying responses to \discrimination \rev{}{,} such as doubling down on studying versus not attending classes\rev{}{,} can be explained by different configurations of appraisals and emotions. Different levels of distress then follow the behavioral response; if it is effective, distress subsides. 

The specifics of CAB responding are influenced by the individual's prior experiences as well as their personal and social resources. These factors are captured within the stress processing model. For example, history of one's exposures and interactions influence the appraisal of new interactions. Similarly, self-regulation skills, developed over time, influence behavioral responding. Thus by combining the stress process and CAB models, we can explain the relationship between experiences of discrimination, short-term response, and long-term outcomes, accounting for the factors that govern the situational appraisal of each experience. 
This joint model in turn has guided our study, which directly explores these relationships using a combination of qualitative and quantitative data. \rev{}{While the general elements of this extended stress process model are already stipulated in the literature, their specifics in the context of student experiences of \discrimination\ are unknown. For example, specifics of the cognitive appraisal in this context is unclear. Our data will help clarify these specifics.}

%% file: 3-method.tex
\section{Methodology for Understanding Experiences of Perceived Discrimination}
\label{sec:study}

We used qualitative and quantitative studies to characterize experiences of \discrimination\ among students. Interviews were used to collect rich insights into how different circumstances and coping behaviors interact to change the trajectory of a person's emotional response over time. Experience sampling data was used to evaluate interview themes in a larger sample. 
Our study was approved by our institution's Institutional Review Board. Following a description of our measure of \discrimination\ and study timeline, we provide details of both experience sampling and interview methods (\sect{sec:study-ema-procedure} and \ref{sec:study-interview}). We then present our findings (\sect{sec:result-qual} and \ref{sec:study-ema}).

\paragraph{Primary Measure of Perceived Discrimination}
Both experience sampling and interviews asked about incidents where participants felt `unfairly treated'. To ensure relevance to discrimination, we eliminated responses that did not connect the unfair treatment to a group affiliation (see \tbl{tab:ema-info} for a list of groups we considered). The choice of wording followed the recommendation for the general study of daily incidents of \discrimination\ \cite{Williams:2009}. Directly using the word `discrimination' leads to under-reporting of everyday cases as the word is commonly used in reference to major incidents \cite{Grollman:2017}. Our data thus captured a range of experiences including microaggressions, everyday discrimination, and other forms of differential treatment. 
While there are important differences among these experiences (\eg in terms of underlying social structures), studies such as ours that look at the  generic psychological response to differential treatment in the short-term may validly include a wide variety of discriminatory experiences. 
Prior literature reports similar responses \cite{Williams:2012} and underlying psychophysiological processes across experiences \cite{Smart:2010} (see \sect{sec:back-microaggression-impact} for details).

\paragraph{Timeline}
We collected reports of \textit{unfair treatment} events from \numuwii students using experience sampling in the 10-week period of Apr 01--Jun 07, 2019. During Jun 10-18, 2019, we interviewed \numinterviewees of the \numdiscrimination individuals who reported discriminatory incidents in the ten-week experience sampling period. Conducting experience sampling before interviews was important for two reasons: 1) it enabled recruitment of students with recent experiences, 2) it provided complementary quantitative data over the same time period of experiences discussed in interviews.

\subsection{Experience Sampling}
\label{sec:study-ema-procedure}

\begin{table*}[bh]
\small
\begin{tabular}{|c|p{25mm}p{11cm}|}

\hline
\parbox[t]{2mm}{\multirow{6}{*}{\rotatebox[origin=c]{90}{\centering \textbf{Affect}}}} & \multicolumn{2}{l|}{\textit{How are you feeling right now? 0(Not at All)-4(Extremely)}}  \\ 
    & \hspace{3mm}Negative & Anxious, Scared, Depressed, Frustrated, Lonely \\
    & \hspace{3mm}Positive & Interested, Enthusiastic, Determined, Inspired, Strong \\
\cline{2-3}
&\multicolumn{2}{p{13cm}|}{\textit{To what extent do you feel like... 0(Never)-4(Very Often)}} \\ 
& \multicolumn{2}{p{13cm}|}{\hspace{3mm}your demands exceeded your ability to cope with them?} \\
& \multicolumn{2}{p{13cm}|}{\hspace{3mm}you are unable to control the important things in your life?} \\
\hline
\parbox[t]{2mm}{\multirow{4}{*}{\rotatebox[origin=c]{90}{\parbox[c]{22mm}{\centering \textbf{Stress}}}}} 
    & \multicolumn{2}{p{13cm}|}{\textit{Which of the following describes the kinds of demands you are currently experiencing? (select all applicable)
    }} \\ 
    & \hspace{3mm}Immediate Peril & interpersonal conflict, criticism or judgement from others, academic concerns (workload, performance, major, exams, projects, homework, \etc), rejection \\
    & \hspace{3mm}Incompetence & feeling incompetent (not doing well in something, failing to meet goals/complete tasks, \etc) \\
    & \hspace{3mm}Others & family concerns, financial concerns, health concerns, personal safety, sexual harassment, barriers in doing tasks (e.g. broken laptop, bad wifi, late bus, lost keys, etc.)
    \\
\hline
\parbox[t]{2mm}{\multirow{2}{*}{\rotatebox[origin=c]{90}{\parbox[c]{24mm}{\centering \textbf{Discrimination}}}}} & \multicolumn{2}{p{14cm}|}{\textit{Did you experience unfair treatment for any of the following reasons today? (select all applicable)} ancestry or national origins, race, gender, sexual orientation, intelligence, major, a learning disability, education or income level, age, religion, a physical disability, height, weight, some other aspect of physical appearance, point of view \vspace{5mm}
} \\ \cline{2-3} 
    & \multicolumn{2}{p{13cm}|}{\textit{How stressful was this experience for you? 1(Not at all stressful)-4(Very stressful)} \vspace{5mm}} \\
\hline
\end{tabular}
\caption[Ecological Momentary Assessment Questions]{Ecological Momentary Assessment (EMA) questions. The category labels (\eg \textit{Negative / Positive} under affect or \textit{Immediate Peril} under stress) were not part of the query participants received. They are added for clarity regarding our research questions. 
}
\label{tab:ema-info}
\end{table*}

We adopted a similar protocol to \citet{Sefidgar:2019}'s 2018 study for our experience sampling surveys: participants were signaled on their phone and via email at least twice weekly to answer Ecological Momentary Assessment (EMA) questions. This sampling regime could be sustained over time while also covering a good subsample of both weekend and weekday experiences. Surveys were hosted on Qualtrics (\url{https://www.qualtrics.com}).

EMA questions included current affect, demands, and experiences of unfair treatment (\tbl{tab:ema-info}). We listed a number of common attributions for unfair treatment (\eg race, gender, socioeconomic status) in the EMA question asking about unfair treatment. We also offered an `other' option with a space for participants to describe the situation following literature best practices \cite{Williams:2008}. This addition was an attempt to mitigate the attributional ambiguity of \discrimination. As \citet{Williams:2009} note targets ``are often uncertain of the reason (or attribution) for a specific interpersonal incident.'' This is particularly true close to the incident because sense-making process is still ongoing. However, the addition of `other' option also allowed reports of unfairness with merely personal nature. We carefully reviewed participant descriptions of such reports and removed six incidents that were based on personal matters (\eg conflicts with significant others) rather than some group affiliation.

Analysis of the experience sampling data was designed to evaluate interview themes. We therefore detail this analysis in \sect{sec:study-ema} and after presenting interview findings in \sect{sec:result-qual}.

\subsection{Interviews}
\label{sec:study-interview}
Each interview lasted about an hour, and included at least two researchers from both psychology and computer science. We tried to create a safe space for participants, by reassuring them that we only cared about their interpretations and feelings rather than judging  what had happened. If participants became upset when describing their experience, we showed support by acknowledging and validating the pain they had experienced. We reminded them that participation was voluntary and they could end the session. We also offered information on counseling and other campus resources. 

\begin{figure*}
    \centering
    \includegraphics[width=\textwidth]{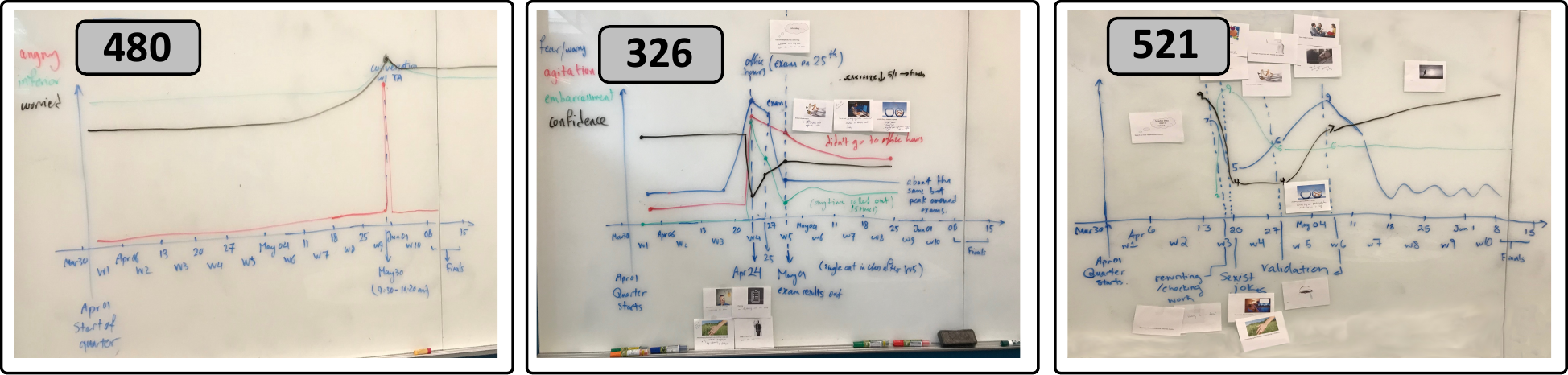}
    \caption[Trajectories of Emotional Response to Perceived Discrimination]{Sample participant drawings. Horizontal axis is week of quarter; Vertical axis shows the intensity of a feeling. These drawings show three different emotional trajectory types. (left) \TrajecoryContinued{} shows a trajectory \rev{}{extreme extended distress}. The participant described feeling worried (black line) and inferior (green line) all quarter, and more so after the incident. The incident (marked with a dotted blue vertical line) caused a brief period of anger (red line) and heightened the other two emotions. (middle) \ResolvedDirectNotConfidence{} described a trajectory that resolved within a few days. (right) \ResolvedIndirectPosTONeg{} had a more complex trajectory with several pertinent events (marked with dotted vertical lines) and partial resolution. \ResolvedIndirectPosTONeg{} described mixed emotions, with an increase in confidence (black line) following distress. This participant used a number of coping strategies (visible on the diagram as pieces of paper).}
    \label{fig:emotion-trajectory}
\end{figure*}

%\paragraph{Perceptions and Appraisals}
%\label{sec:study-interview-procedure-perceptions}
\subsubsection{Interview Protocol} 
\label{sec:study-interview-protocol}
The interview was structured around the CAB framework (blue arrow in Figure~\ref{fig:microaggression-model-stress}). We first explored participants' \textit{appraisal} of what had happened: 
we asked participants to describe in as much detail as possible one incident of \discrimination\ during the ten-week EMA period that stood out the most. We followed up with questions to understand the meaning and significance of the event and the surrounding circumstances as well as changes in interpretations over time. 
%\paragraph{Emotional Response}
%\label{sec:study-interview-procedure-emotions}
We also asked participants to review their messages and calls around the time of the incident to spur recall (``scroll-back'' \cite{Robards:2017}).
Next, we asked about participants' \textit{affective responding} (emotional reactions). Following an open-ended prompt to avoid priming participants, we asked about feelings commonly associated with experiences of discrimination \cite{Carter:2010, Carter:2013}. These included helplessness, hopelessness, or sadness (depression); fear, agitation, nervousness, or worry (anxiety); empowerment, motivation, or confidence (vigor); rejection, inferiority, or isolation (self-esteem); shame or guilt; and anger. We asked participants to graph their \textit{emotion trajectories}, for three or four feelings they thought most strongly represented their experience, on a whiteboard (\fig{fig:emotion-trajectory}). The horizontal axis is \begin{wrapfigure}{R}{0.4\textwidth}
\vspace{-1.5em}
\includegraphics[width=0.4\textwidth]{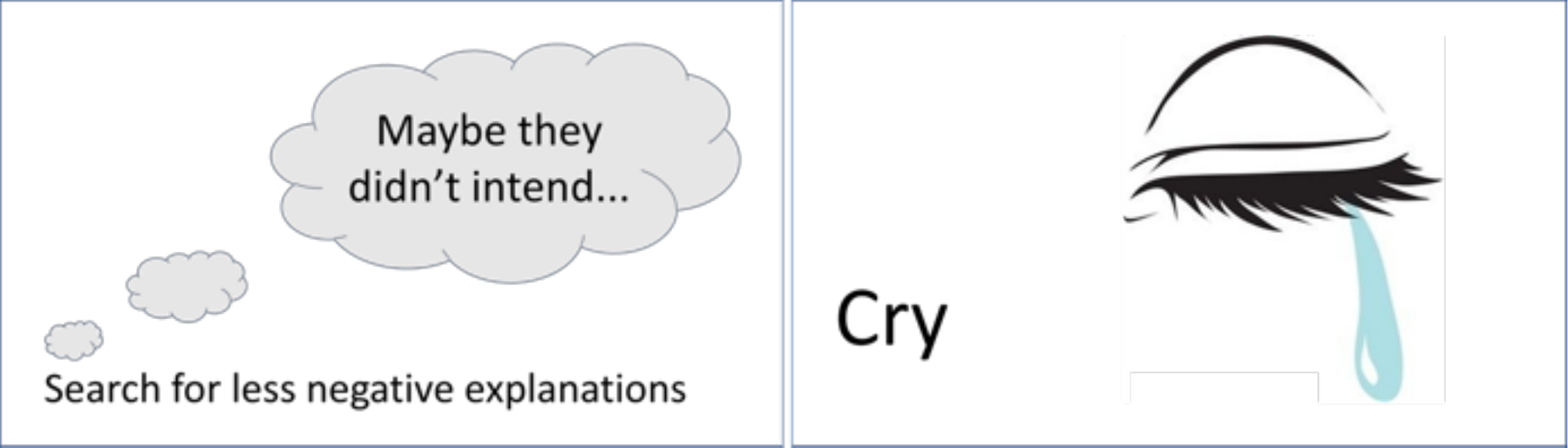}
    \caption[Sample Coping Cards]{Sample coping card. Each card described a coping strategy in words and images. 
    } 
    \label{fig:coping-cards}
%\vspace{-1em}
\end{wrapfigure}

\noindent weeks of the quarter and the vertical axis is the strength of emotions on a scale between 0 (none) and 10 (the strongest). Finally, we asked participants about how they \textit{coped} with the incident. We gave them \textit{coping cards} to comment on based on relevance (\fig{fig:coping-cards}). Coping cards were based on the literature on stress and coping \cite{Carver:1997}, emotion regulation \cite{Gross:2015}, and discrimination \cite{Carter:2010, Forsyth:2014} and included about 30 different common ways of coping, such as crying, exercising, yoga, planning, shopping, ruminating, mindless scrolling and self-blaming (See Appendix). Relevant cards were placed along the emotional trajectory according to when they were used (\fig{fig:emotion-trajectory}, middle and right). 

\subsubsection{Interview Analysis}
\label{sec:analysis-qual}

Interview sessions were followed by interpretation sessions where researchers discussed the interview. Each interview session was audio recorded and transcribed. Data from each interview \rev{}{additionally} included notes taken during the interview, pictures of emotion trajectories and coping cards, and observations by the interdisciplinary team members present in each interview. Analysis began immediately after the first interview and was iterative. %\paragraph{Researcher Relationship to Topic}
%\label{sec:study-interview-procedure-reflexivity}
Study team members were of different ethnicities, academic disciplines, and career stages. They, at times, related in different ways to participant accounts of discriminatory incidents. During debrief sessions, researchers openly discussed their subjective reactions and related observations and questions with each other.

Our analysis addressed two questions: (1) what are the prominent response patterns? (2) what factors influence response, thus leading to distinct patterns? Our analysis included a close review of all collected data, notes taken during the interviews, and notes taken during discussions among research team members. We used a combination of top-down and bottom-up coding, adhering to principles from thematic analysis  \cite{Braun:2006}. Top-down codes primarily categorized coping strategies based on existing psychology literature. We also coded for emergent bottom-up codes including semantic (\eg individuals stated whether or not they had engaged in self-blame) and latent concepts (\eg we inferred internalized biases when a participant commented ``I didn't want to be a bitch by confronting them'', when describing a gender-based discriminatory treatment).  
As themes emerged from this process, they were reexamined and critiqued with a larger research team, who reviewed both raw interview material and interview summaries. 

\subsection{Participants}
\label{sec:study-participants}
We recruited first and second year college students to participate in the experience sampling survey. We advertised across the university, and also used targeted advertising to reach people from a variety of minoritized identities who were more likely to experience unfair treatment. The interview sample was drawn from the subsample of experience sampling participants who reported unfair treatment during the ten-week experience sampling. 

%\subsection{Participants}
%\label{sec:study-ema-participants}
\tbl{tab:uwexpii-info} summarizes demographics of the participants in experience sampling and interviews. 
Our \numuwii experience sampling participants included 174 people who self-identified as Asian, Black, Latinx, or Biracial, of whom 58 reported  unfair treatment, and 127 people who identified their gender as non-male, of whom 62 reported unfair treatment. In addition, 7 of 25 non-heterosexual students, and 9 of 23 people who identified as disabled reported unfair treatment. 

\begin{table}[h]
\centering
\begin{tabular}{rc|c|c}
\multicolumn{1}{l}{}   & \textbf{\begin{tabular}[c]{@{}c@{}}Experience Sampling\\ (N = \numuwii )\end{tabular}} & \textbf{\begin{tabular}[c]{@{}c@{}}Reported Unfair Treatment\\ (N = \numdiscrimination)\end{tabular}} & \textbf{\begin{tabular}[c]{@{}c@{}}Interviewed\\ (N = \numinterviewees)\end{tabular}} \\ \cline{2-4} 
People of color        & 174 (68.8\%) & 58 (64.4\%)  & 10 (71.4\%) \\
\rev{}{Women and other gender minorities} & 127 (50.4\%) & 62 (68.9\%)  & 8 (57.1\%) \\
\rev{}{Sexual orientation minorities} & 25 (9.9\%) & 7 (7.8\%)  & 1 (7.1\%) \\
People with disability & 23 (9.1\%) & 9 (10.0\%)  & 1 (7.1\%)                
\end{tabular}
\caption[Demographics of Experience Sampling and Interview Studies]{Participant demographics. \% in a column is based on N in that column. People of color self-identified as Asian, Black, Latinx, or Biracial. \rev{}{Gender minorities included transgender, genderqueer, and others who did not associate with any of these categories. Sexual orientation minorities included homosexual, bisexual, asexual and others who did not associate with any of these categories.} People with disability self-identified sensory, mobility, learning, or cognitive disability.}
\label{tab:uwexpii-info}
\end{table}

Table~\ref{tab:participant-info} summarizes the incidents described by the \numinterviewees interview participants. The interviewees were representative of the subsample of experience sampling who reported \discrimination: we tested demographic characteristics that influence stress exposure 
(sex and race only given the small sizes of other categories) and cumulative stress of prior exposure to \discrimination\ (measured by combining Everyday Discrimination \cite{Williams:1997} and Chronic Work Discrimination and Harassment \cite{Williams:2008} scales, EDSCEDH). We also tested for difference in general coping behaviors based on student responses to BriefCOPE scale \cite{Carver:1997}. We applied $\chi^2$-test for comparing categorical variables (sex and race)  and t-test for comparing numeric ones (EDSCEDH, and adapative and maladaptive subscales of BriefCOPE). All comparisons were performed in R \cite{R}. No significant differences were found on any of these characteristics.

\begin{table*}[]
\centering
\resizebox{\textwidth}{!}{%
\begin{tabular}{c|c|p{1.3cm}|p{2.2cm}|p{10cm}|p{1.8cm}|c|}
\cline{2-7}
\textbf{} & \multicolumn{3}{c|}{\textbf{Demographics}} & \multicolumn{3}{c|}{\textbf{Perceived Discrimination Events Discussed in the Interview}} \\ \cline{2-7} 
\textbf{PID} & \textbf{G} & \textbf{Ethnicity} & \textbf{Intended Major} & 
\textbf{Perceptions of What Happened} & \textbf{Direct Peril} & \multicolumn{1}{l|}{\textbf{Resolved}} \\ \hline
\multicolumn{1}{|c|}{307} & M & Asian & Engineering & 
A professor dismissed his learning needs and instead used his poor performance in a single test as an indicator of low intelligence and lack of fit for Engineering. & Not getting into major & N \\ \hline
\multicolumn{1}{|c|}{326} & M & Asian & Biology & 
A professor racially targeted him by singling him out and posing a difficult question as he tried to leave office hours. & Receiving low grade & Y \\ \hline
\multicolumn{1}{|c|}{362} & M & BIPOC & Engineering & 
An advisor disparaged his efforts and questioned his desire for learning based on his socioeconomic background. & Being ousted from program & N \\ \hline
\multicolumn{1}{|c|}{404} & W & Asian & Engineering & 
Male classmates laughed at a question she asked in class. &  & N \\ \hline
\multicolumn{1}{|c|}{411} & W & BIPOC & Engineering & 
A peer told her that she would not be admitted to an engineering major given her socioeconomically disadvantaged background. & Not getting into \mbox{major} &  N \\ \hline
\multicolumn{1}{|c|}{420} & M & Asian & Engineering & 
A bouncer at a fraternity party mocked his appearance, saying that he, an Asian man, looked too young. &  & N \\ \hline
\multicolumn{1}{|c|}{424} & NB & White & Engineering & 
An online survey did not offer a proper box for gender. &  & N \\ \hline
\multicolumn{1}{|c|}{471} & M & BIPOC & Engineering & 
A professor ignored him as he was trying to communicate his question as a non-native English speaker. & Receiving low grade & Y \\ \hline
\multicolumn{1}{|c|}{474} & W & BIPOC & Engineering & 
Peers criticized her for her religious views. & Being socially excluded & N \\ \hline
\multicolumn{1}{|c|}{480} & W & Asian & Engineering & 
A male TA dismissed her complaints that a test was based on material not covered in class, telling her she was unprepared. & Receiving low grade & N \\ \hline
\multicolumn{1}{|c|}{488} & W & BIPOC & Public Health & 
A TA racially profiled her and accused her of plagiarism without solid basis. & Facing academic penalty & Y \\ \hline
\multicolumn{1}{|c|}{521} & W & White & Engineering & 
Male peers took credit and explained her own work to her. &  & Y \\ \hline
\multicolumn{1}{|c|}{527} & W & Asian & Engineering & 
Male peers accused her of coming to study sessions only to get answers. &  & N \\ \hline
\multicolumn{1}{|c|}{552} & W & White & Neurobiology & 
A male peer mocked her as a ``dumb blond'' as they worked on a class exercise. &  & N \\ \hline
\end{tabular}%
}
\caption[Interviewee Demographics and Reports of Discrimination]{Interviewee demographics and \discrimination\ events they described in the interview. \textit{G} (Gender categories) are men (M), women (W), and non-binary (NB). Categories of ethnicity are BIPOC (Black, Latinx, or Biracial), Asian (Middle-eastern, South-Asian, and East-Asian), and White. The representation of ethnicity with high level information is intended to protect participant identities. \textit{Resolved} is N (no, not resolved) or Y (yes, resolved).
\textit{What Happened?} briefly summarizes the incident as perceived by participants.
We note that these short descriptions do not capture the full complexity of the experiences. Nonetheless, we hope to convey that the key aspect of all the incidents is that they are perceived as behaviors stemming from negative attitudes and treatments toward a group of people. They are not appraisals of unfairness of merely personal nature.
}
\label{tab:participant-info}
\end{table*}

%% file: 5-result.tex
\section{Findings from Interviews}
\label{sec:result-qual}

We begin by providing an overview of the \textit{most significant emotional response trajectories} described by participants. We then detail the three categories of factors that influenced these trajectories: 
underlying `perils' that amplify distress; beliefs that intensify distress and complicate response; and the timing and strategy of coping that impact outcomes.

\subsection{Distinct Intensity, Onset, Duration, and Quality of Emotions Characterize Response Trajectories}
\label{sec:result-qual-emotion}

\begin{figure}
\centering
\begin{minipage}{.4\textwidth}
  \centering
  \includegraphics[width=\linewidth]{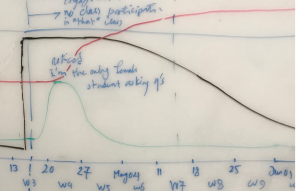}
  \captionof{figure}{\NResolvedIndirectDelay. Dotted vertical line on the left: \discrimination\ event; trajectories in blue (isolation), red (motivation), green (sadness), and black (embarrassment).}
  \label{fig:trajectory-404}
\end{minipage}%
\hfill
\begin{minipage}{.5\textwidth}
  \centering
  \includegraphics[width=0.8\linewidth]{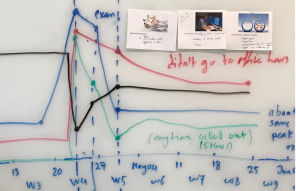}
  \captionof{figure}{\ResolvedDirectNotConfidence. Vertical lines from left to right: \discrimination\ event, exam (difficult situation, the ``peril''), and exam results (resolution of ``peril''); trajectories in blue (fear), red (agitation), green (embarrassment), and black (confidence).}
  \label{fig:trajectory-326}
\end{minipage}
\end{figure}

% types of emotions \& curves
Not surprisingly, students described negative feelings such as sadness, isolation, anxiety and shame/embarrassment, referred to as \textit{distress} hereafter, in response to \discrimination. As highlighted in participants' whiteboard drawings, distress varied in intensity and in length from a few hours to several weeks (\fig{fig:emotion-trajectory}). Some negative feelings immediately followed the incident, while others were delayed. For example, sadness usually peaked with some delay and after continued appraisal of the event. \NResolvedIndirectDelay, a gender minority who had been laughed at in class after asking a question, became increasingly upset (\fig{fig:trajectory-404} - green line) even as the initial embarrassment (black line) began to fade. It was slightly later that the career implications became apparent \qt{What have I got myself into?} she asked, about a department and field where women are under-represented. %\yasaman{NOTE 471 as another example}

Students reported increased confidence when they resolved the situation, although it did not always rebound to pre-event levels. For example, \ResolvedDirectNotConfidence's  confidence (\fig{fig:trajectory-326} - black line) dropped significantly after he felt racially targeted by an instructor who challenged his grasp of the material, and only partially recovered after receiving a good score on the course exam (third dotted blue line).

\begin{figure}[h]
    \centering
    \includegraphics[width=.7\textwidth]{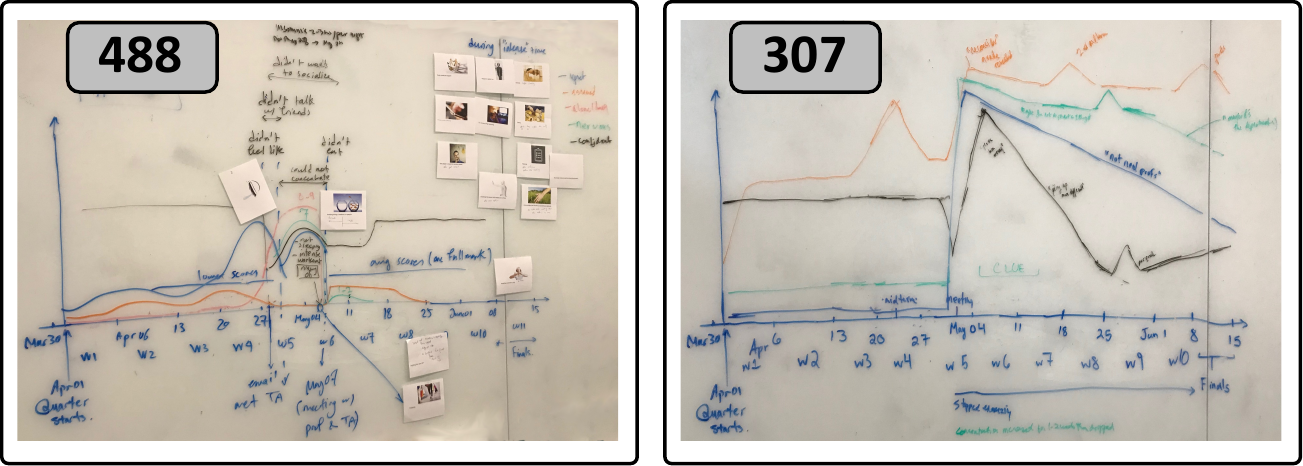}
    \caption{(Left) \ResolvedDirectPosTONeg's confidence (black) increased as her anxiety (green) and shame (orange) dropped. (Right) \NResolvedDirectNegTOPos's motivation (black) decreased as inferiority (green) and nervousness (orange) increased.}
    \label{fig:trajectory-488-307}
\end{figure}

% inflection points
Several participants described adaptive transitions between emotions (\eg a shift from sadness to motivation), while others described the opposite (\eg a temporary surge in motivation followed by discouragement and disengagement from a class). An adaptive transition played out as a woman, who had previously felt hurt by sexist comments of peers, corralled administrators and students to discuss bias; \ResolvedIndirectPosTONeg's confidence (black) increased as her frustration (blue) decreased, \fig{fig:emotion-trajectory} -- right. Transitions to positive emotions also occurred when students changed study habits and saw evidence of their efforts in the form of higher grades (\ResolvedDirectPosTONeg, \fig{fig:trajectory-488-307}).
In an opposite case, a student who was initially motivated to disprove a professor's doubts in his intellectual ability, gradually lost motivation as he continued to struggle in the class (\NResolvedDirectNegTOPos, \fig{fig:trajectory-488-307}). 

\subsection{Underlying `Perils' Amplify Distress}
\label{sec:result-qual-peril}
As we compared emotional responses across participants, we were struck by a differentiating factor: students who had the strongest reactions and took longer to recover were those who faced academic or social `perils': underlying struggles, challenges, or vulnerabilities that threatened academic or social standing. \Discrimination\ became entangled with the peril, making a bad situation even worse (in addition to being intrinsically unfair).
Perils posed different levels of immediacy; the possibility of failing a class was weeks away, whereas the possibility of encountering gender inequity in the workplace was, for a college student, further in the future. 
Immediate perils were especially distressing, \eg when \discrimination\ related to the risk of failing a class (\DirectClass), losing one's status in a program (\DirectProgram), or being denied access into one's intended major (\DirectMajor). 
%\subsubsection{Distress Follows the Primary Appraisal of Peril}
\label{sec:result-qual-peril-distress-primary}
For example, 
\ResolvedDirect{}'s question was dismissed during office hours a few days before the midterm exam. He was struggling with the material and sought help from the instructor. Speaking English as a second language, he had difficulty communicating his question. Instead of attempting to understand the question, the instructor ignored him and moved on to answer other students. He was so distressed that he considered dropping the class.

Distant perils, \eg a sexist classmate who did not threaten one's standing in class but was a reminder of broader gender bias issues that could impinge on the individual's future career advancement (\IndirectGender), caused less intense distress. 
Distress was briefest for students who experienced \discrimination\ with no significant peril. For example, \ResolvedIndirectInsignificant, an Asian student, was mocked by a bouncer at a fraternity party, which posed no threat to his major goals. His distress was short-lived: \qt{I mean, it's a party, so at the end of the day we were all trying to have fun.}

When the assessment of perils changed over time, so did distress. This appeared independent of immediacy of the associated perils. For example, both \NResolvedIndirectDelay{}'s distress when she was laughed at (no immediate peril) and \ResolvedDirectDelay{}'s  distress when his professor dismissed his questions before an exam (immediate peril) increased over time as they reflected on the potential consequences for their careers. Relatedly, when a peril resolved, distress decreased. For example, \ResolvedDirect's distress considerably decreased after clarifying the class material and getting a good grade: \qt{... I got my results and I did well... the fear went down...} When the peril was distant (\eg concern about gender bias within Science, Technology, Engineering, and Math (STEM) that would affect one's career), recovery from the emotional distress was sometimes, but not always, possible without addressing the peril, \ie the larger social challenge. 

\subsection{Beliefs Intensify Distress and Complicate Response}
\label{sec:result-qual-belief}

Existing insecurities about one's belonging and efficacy intensified and prolonged reactions, particularly in women. We found that female student coping was limited by competing beliefs about expected feminine behavior as well as some of the institutional messaging in STEM programs, where women are considerably underrepresented. 

Consistent with literature (\eg \cite{Becerra:2020}), \discrimination\ threatened self-esteem and beliefs of belonging, particularly for women. In many cases, participants \discrimination\ as messages that they were incapable and unlikely to succeed.

\rev{}{We observed that p}articipants who doubted their ability (these participants were all women) 
For example, \LowCapacityIntellect{} felt at a disadvantage in her competitive program: \qt{People are like, `Oh, there's tier 1, tier 2, tier 3,' and they're like, `You know, you're really good at studying hard, but you're not the best.'} She described her TA's criticism of her 
preparation as \qt{He was attacking my intellect.}, 
adding that \qt{The TA already putting that [the criticism] on top of that [not seen as tier 1], it makes me just feel like, Okay, maybe.} The distress she experienced was so strong that later on the day of the event, \LowCapacityIntellect{} broke down crying in the presence of an academic advisor. The advisor walked her to the counseling center. 
Attributing blame to herself rather than external causes she could change \LowCapacityIntellect{} did not resolve the situation. 
Students without these beliefs could see negative circumstances as changeable and act constructively to change them.  
For example, \HighCapacity{} did not question his capability to be successful after he was unable to communicate with his instructor. He was worried about dealing with the class but wondered \qt{How can I improve the way I'm speaking?} instead of doubting his communication efficacy. 
Moreover, \HighCapacity{} attributed blame to an external cause. He realized, in the course of successfully conveying his question to a peer, that the problem was the professor's attitude. \HighCapacity's belief in his efficacy was integral to his resolution of the challenges.

We noticed different self-efficacy beliefs 
were activated depending on the nature of the \discrimination\ event. Gender-related \discrimination\ by authority figures (\ie professor, TA) in the context of school performance more directly triggered beliefs of inefficacy and were associated with greater distress than those by peers in our sample. For example, \LowOutcomeEfficacy{} could not dismiss her TA's comment on her performance: \qt{Small comments like that already impact me so much because I already have this feeling of being inferior to my male classmates even though I'm performing just as well as them. And, when he [TA] said that comment, and he has a degree, and he told me that...} She felt her TA was in a position to accurately appraise her intellect, which generated greater distress. In contrast, \NeutralOutcomeEfficacy{} could more easily distant herself from her peer's comments on her intellectual ability: \qt{He doesn't know me, he doesn't know anything about me ... I definitely know it wasn't the end of the world.
} \NeutralOutcomeEfficacy's perceptions of her intellectual efficacy were not strongly influenced by the comments that came from a peer who did not know her.

Beliefs about norms impacted the coping strategies students brought to bear. For example in the face of \discrimination\ associated with social norms, only \NoSocialConcern{} advocated for change 
by raising awareness about bias. One reason was concerns about social rejection. For example, \SocialConcernEGii{} described: \qt{Whenever a girl is ambitious, it comes off as bitchy.} When explaining her response to peers who accused her of abusing study sessions, she added \qt{There is this silent Asian girl in my cohort, and I just feel like maybe I should be more like her, ... `Don't express your opinions, like her, and then you'll be fine.}.  Institutional messaging oblivious to these social inequities further undermined coping with gender discrimination (\SilentCode). \SilentCodeiii{} described how her cohort was encouraged to form a strong community: \qt{[We] are supposed to help each other. We're supposed to study together ... They're my friends.} She added this emphasis on community prevented her from advocating for herself and confronting sexist statements and behaviors. The other two women who were also in her cohort shared the same concerns.

\subsection{Coping: It's Not Just What, but When and How}
\label{sec:result-qual-coping}

Drawing from stress and coping literature \cite{folkman2011oxford}, we defined meaningful recovery as both regulating feelings and resolving the underlying perils. Expanding the literature on the importance of active and problem-focused coping for \discrimination\ \cite{Schmitt:2014}, we observed that meaningful recovery from distress depended on applying a diverse repertoire of these strategies  %(\tab{fig:resolution-coping}---rows 27-33) 
and on doing so persistently and strategically. Moreover, some coping strategies (\eg crying) offered limited adaptive value. However, they impeded recovery if there was no transition to active and problem-focused coping. Consistent with the literature, we saw certain coping strategies (\eg self-blame) persistently hindered recovery \cite{Valentiner:1996}. 

Also consistent with the literature \cite{Craig:2014, Mossakowski:2014}, effective social support facilitated meaningful recovery. Prior work has demonstrated the source of social support impacts its effectiveness in the context of family and friends \cite{Mossakowski:2014}. We found similar effects in an academic context. Specifically, `seeking upward-focused help or company', in the form of inspiration, role modeling, or guidance from someone with relevant knowledge or experience (\eg a fellow student who knew the material or had handled a certain challenge before), was more effective than `commiseration', defined as confirmation of one's misery. For example, \ResolvedUpwardHelpi{}, whose situation resolved, reached out to a peer to get his question answered after his instructor dismissed him, whereas \NResolvedUpwardHelpii{}, whose situation did not resolve, looked to an academic advisor to sympathize with his struggles rather than seeking advice on more effective study strategies. `Commiseration'  %(\tab{fig:resolution-coping}---row 11)
was more prevalent among those who did not meaningfully \rev{}{resolve the incident}. Six of \numintervieweesnresolved participants who did not meaningfully \rev{}{resolve} commiserated (\NResolvedCommiserate) whereas only one who meaningfully recovered did so (\ResolvedCommiserate). Following up on commiseration with other forms of support seemed to be the key. For example, \ResolvedCommiserate{} sought upward support after initial commiseration. Students who commiserated but did not resolve their struggles did not reach out for other support after commiseration. Take, for example, \NResolvedCommiserateEGii, who was told by her peer that she would not get into her intended major. She sought comfort from a friend who was told the same thing by the peer; she did not seek or receive guidance about how to increase her chances of getting into her desired major. 

Many coping strategies that are categorized in the literature as unproductive were only problematic in some circumstances. For example, distraction techniques, such as scrolling, binge watching, and game playing,  %(\tab{fig:resolution-coping}---rows 14,17/18), 
were sometimes brief emotion regulation strategies that helped an individual pause before re-engaging in a challenging situation. An example is \ResolvedADistraction{} who watched Netflix on the evening of the discriminatory event before going into a more intensive study effort. However in many other cases, distraction strategies functioned as avoidance without actively addressing the \discrimination. For instance, \NResolvedADistraction's involvement in an extra-curricular club diminished time and energy that he could have dedicated to the class in which he was struggling. The club provided emotional support, but it was also an ongoing distraction from improving his standing in the class. Similarly, avoiding people or places 
%(\tab{fig:resolution-coping}---24) 
was occasionally helpful. For example, \AvoidancePosi{} avoided the peer who told her she could not become an engineer. However, avoiding people or places was sometimes a barrier to success and particularly problematic if not followed by active and problem-focused coping. \AvoidanceNegii{} avoided class without engaging in more active and problem-focused coping. She became more isolated, and her beliefs about not belonging increased. Isolation contributed to her distress and hopelessness. To summarize, \textit{how} and \textit{when} different coping strategies were used impacted recovery, as did the \textit{diversity} of those strategies and their combination with each other.

%% file: 6-quant.tex
\section{Evaluating Interview Observations with Experience Sampling Data}
\label{sec:study-ema}

Based on our interviews, we identified two research questions that were valuable and feasible to test in the larger experience sampling data. They provided confirmatory support for the themes extracted in our qualitative analysis.

%\subsection{Research Questions}
\label{sec:study-ema-qs}

\begin{enumerate}[start=1,label={\bfseries RQ\arabic*}, leftmargin=1.5cm]
    \item \label{itm:rq-distress} Our interviews suggest that distress is frequently highest shortly after incidents of \discrimination. Thus, we ask: \textit{Are there differences in self-reported daily affect when students report unfair treatment? }
    \item \label{itm:rq-confirmation} Our interviews showed that underlying perils and beliefs influence the intensity and duration of reactions. Thus, we ask: \textit{Are there differences in psychological distress reported for unfair treatment in the presence versus absence of (a) immediate perils, (b) beliefs of incompetence?}
\end{enumerate}

%\subsection{Results of Confirmatory Analysis of Experience Sampling Data}
\label{sec:study-ema-results}
\paragraph{RQ1: Analysis of Distress Associated with Perceived Discrimination}
\label{sec:analysis-quant-rq2}
Interviewees described negative feelings that immediately followed incidents of \discrimination. We examined this observation in experience sampling data by comparing self-reported affect in the presence and absence of incidents. We derived indicators of positive and negative affect by averaging responses to [\textit{interest, enthusiasm, determination, inspiration,} and \textit{strength}] and [\textit{anxiety, fear, depression, frustration,} and \textit{loneliness}], respectively. We defined feeling overwhelmed as the average of responses to questions on perceived levels of demands and control over them (\tbl{tab:ema-info}). We constructed two-level hierarchical linear models (HLMs) for each variable to capture the relation between affect (outcome variables) and unfair treatment events (input variable). HLMs can handle unequal number of unevenly-spaced reports per person. We set up the models to compare outcomes for each participant on days they reported events and days they did not. We used \textit{lme} from \textit{nlme} package \cite{R:nlme} to create three HLMs for outcome variables negative affect, positive affect, and feeling overwhelmed.

We found a significant increase in negative affect on days unfair treatment was reported (p < .001). This is consistent with interview findings as well as past literature \cite{Ong:2009, Sefidgar:2019}. Moreover, we found no evidence for significant associations between unfair treatment reports and positive affect which is consistent with literature \cite{Potter:2019}. Our analysis did not produce evidence for the relations between reports of unfair treatment and feelings of being overwhelmed. This is likely because the appraisal process was still ongoing at the time students were signaled to complete EMA surveys. \tbl{tab:quant-result-rq2} details the results.

\paragraph{RQ2: Analysis of Perceived Discrimination Distress under Immediate Perils and Beliefs of Incompetence}
\label{sec:analysis-quant-rq3}
In interviews, we observed intensified distress associated with \discrimination\ linked to immediate perils or beliefs of incompetence. We operationalized \textit{Immediate Perils [$0,1$]:} $1$ if there were any reports of academic concerns (workload, performance, major, exams, projects, homework, \etc), rejection, interpersonal conflict, and criticism or judgment from others. We operationalized \textit{Incompetence [$0,1$]:}  $1$ if there were any reports of feeling incompetent (not doing well in something, failing to meet goals/complete tasks, \etc). We created an ordinal logistic regression model with distress associated with \discrimination\ as outcome, and immediate peril and incompetence as input. 
In consultation with statistics experts, we chose the ordinal logistic regression model over the linear one as the outcome variable was Likert-style rather than numeric. 
Our model allowed us to compare distress of \discrimination\ in the presence and absence of immediate perils and self-perceived incompetence for each individual and could properly handle unequal number of unevenly spaced reports per individual. We used \textit{clmm} from \textit{ordinal} package \cite{R:ordinal} to create the model. We found that \textbf{students report higher levels of distress for incidents of unfair treatment when they also report immediate perils and they feel incompetent.} The regression coefficient is comparatively stronger for self-perceived incompetence (\tbl{tab:quant-result-rq3}). This supports our interview findings that perils and beliefs can influence the intensity of distress associated with \discrimination.

\begin{table}[t]
\small
\parbox{.45\linewidth}{
\begin{tabular}[t]{|p{28mm}|c|c|c|}
\hline
\textbf{Same-Day} & \textbf{b} & \textbf{SE} & \textbf{p} \\ \hline
Positive affect & -0.06 & 0.07 & 0.42 \\
\textbf{Negative affect} & \textbf{0.42} & \textbf{0.06} & \textbf{< 0.001} \\
Feeling overwhelmed & 0.08 & 0.07 & 0.22 \\ \hline
\end{tabular}
\caption[Relationship between Perceived Discrimination and Daily Affect and Stress]{Relationship between \discrimination\ and daily affect and stress. (b: regression coefficient estimate, SE: standard error of the estimates, p: p-value). Significant relations are in bold. 
}
\label{tab:quant-result-rq2}
}
\hfill
% \hspace{-1cm}
\parbox{.5\linewidth}{
\begin{tabular}[t]{|p{38mm}|c|c|c|}
\hline
\textbf{Discrimination Distress} & \textbf{b} & \textbf{SE} & \textbf{p} \\ \hline
\textbf{Immediate perils} & \textbf{0.52} & \textbf{0.23} & \textbf{0.02} \\
\textbf{Incompetence} & \textbf{0.7} & \textbf{0.23} & \textbf{0.003} \\ \hline
\end{tabular}
\caption[Relationship between Perceived Discrimination Distress and Perils and Beliefs of Incompetence]{Relationship between distress of \discrimination\ and perils and beliefs of incompetence. (b: regression coefficient estimate, SE: standard error of the estimates, p: p-value). Significant relations are in bold.}
\label{tab:quant-result-rq3}
}
\end{table}

%% file: 7-discussion.tex
\section{Discussion}
\label{sec:discussion}
Consistent with other research \cite{Potter:2019}, we found \discrimination\ to be psychologically burdensome for our participants. Our results help untangle how these experiences were shaped by perils, beliefs, and coping behaviors. Most prominent in our qualitative and quantitative results was the strong relation between distress of \discrimination\ and immediate perils. 
This relation was typically made worse by associated beliefs of incompetence, or coping strategies that did not actively address the peril. 
Below, we discuss the findings through the lens of the extended stress processing framework. This theoretical grounding \rev{}{points to} areas we can influence to discount short-term distress and eventually improve long-term outcomes. 
We build up on this stress processing discussion to highlight \rev{}{some of} the relevant technologies and their requirements. 

\subsection{Insights from the Extended Stress Process Model: Beyond In-the-moment Comfort} 
\label{sec:discussion-model}

We use\rev{}{d} the extended stress process model of \sect{sec:back-microaggression-model} \rev{}{as a conceptual framework} to \rev{}{guide our study}. As a reminder: the extended stress process model considers short-term distress that is embedded within longer-term processes and outcomes. Short-term distress is described as a process of {\em C}ognitive, {\em A}ffective, and {\em B}ehavioral (CAB) responding (large blue arrow in \fig{fig:microaggression-model-stress}) and sits within the stress process model (gray box in \fig{fig:microaggression-model-stress}), which explains pathways to long-term outcomes. \rev{}{
While the high level elements of the model (\eg cognitive appraisal) are well established, our work characterizes the specifics of these elements in the context of student experience: our observations around perils, beliefs of incompetence, and patterns of effective coping clarify what cognitive appraisal and behavioral responding processes are prominent in student experience. 
Using the conceptual framework of 
the extended stress process model, we highlight (1) how these factors play a role in shaping not only short-term distress, but also long-term consequences and (2) why they should be addressed in providing meaningful and extended support. 
We emphasize that the importance of these factors and that they matter in mitigating the harm of \discrimination\ was not recognized until this analysis. Our work therefore significantly expands the intervention design space of existing literature, which is primarily focused on emotional comfort needs (\eg \cite{To:2021}).
}

\rev{}{More broadly, our theoretical grounding offers} a disciplined approach to intervention design: \rev{}{if we identify the specifics of CAB processing and build solutions to address them, we can support individuals in managing short-term distress. Such support can also discount long-term consequences. }
The extended stress processing framework further suggests a need to consider\rev{}{, and address,} effects that develop over time \rev{}{as arenas of support against the impact of \discrimination}. For example, prior stress exposure may influence the likelihood that students find themselves in high cost and perilous situations, the beliefs they develop, and their coping resources. \rev{}{Our analysis highlights that} the design space of support technologies in face of \discrimination\ goes far beyond \textit{in-the-moment} help, \rev{}{the primary focus of the current literature, \eg in the context of racist incidents \cite{To:2020, To:2021}}. We should design for the individual as a whole and beyond the individual at the moments of crisis. In case of \discrimination\ in academic settings, this means \rev{}{we need technologies} that make it less likely for students to face struggles, form dysfunctional beliefs, and lack coping resources. 

\rev{}{In t}he next subsections \rev{}{we} expand on \rev{}{this} theoretical grounding \rev{}{and discuss the needs that technology solutions should address. We organize our discussion under two intervention paradigms:} (1) \textit{`incident-specific' intervention paradigm}: solutions that aim at helping individuals \textit{react} to each discriminatory incident, and (2) \textit{`proactive' intervention paradigm}: solutions that cultivate anticipatory preparation to \textit{proactively} address vulnerabilities 
and to build up coping resources to reduce the likelihood that students experience intensified distress in the first place. In addition to their grounding in the stress processing frameworks, these paradigms are supported by the general coping literature \cite{Schwarzer:2008}.

\subsection{Incident-Specific Interventions: Managing Short-term Distress}
\label{sec:discussion-guide-reactive}
Interventions that support students as they react to distress of incidents of  \discrimination\ can be considered under the umbrella of mental health support technologies. We discuss three requirements for these technologies:  assessment of vulnerabilities, holistic monitoring of response, and guided coping.

Interventions should account for vulnerabilities such as perils or problematic beliefs that intensify distress. 
Currently, inquiry about user situation in mobile and internet interventions functions primarily as a means to user engagement~\cite{Fitzpatrick:2017}. 
To address \discrimination, this contextual information could explicitly drive functionality. For example, the opening question from \citet{Fitzpatrick:2017}'s conversational agent Woebot, `What's going on in your world right now?' could be used to adjust intervention to the relational context of a stressful incident (\eg the need to understand course material when differentially dismissed by course staff).
Similarly, it is key to address beliefs of inefficacy or an appraisal of a negative situation as uncontrollable when adapting solutions, such as CBT-based technologies \cite{Hur:2018, Kim:2020}, to the context of \discrimination.
\rev{}{Helping} students recognize and value their abilities \rev{}{is key to effectively resolving the situation}.

Distress associated with \discrimination\ unfolds over time as a situation is reexamined, emotions change, and needs evolve. 
Interventions should therefore assess emotions over time. Further, negative emotions should not necessarily be interpreted as problematic nor positive emotions as progress. For example, negative emotions may accompany persistence at an important goal and positive emotions could reflect mere distraction instead of meaningful coping. Support should therefore account for both emotional states and indicators of effective coping.
Building up on research to monitor emotions \cite{Burns:2011, Mohr:2017-sensing} and adapting to them \cite{Fitzpatrick:2017},
interventions should interpret emotions in tandem with behavioral and academic data. 
The goal is to help people feel good while providing meaningful support. Both passive (\eg sensor-based) and active (\eg conversational) solutions could be relevant. %explored to address this need.

Transitioning to strategic and persistent active coping and upward social support seeking is key for managing the distress that ensues \discrimination.
Solutions should support such transitions and the strategic and persistent application of these skills by bolstering students' resources and scaffolding their effort.
Technologies that provide evidence-based guidance for social support \cite{OLeary:2018, Wadden:2020} (\eg through guided conversations) and active coping (\eg through goal setting, accountability, and reflection modules) \cite{Fitzpatrick:2017, schroeder2018pocket} have shown promise and could be useful in the context of \discrimination. 
We specifically call out the importance of drawing on shared frame of reference in social support systems. Students should be able to connect to those who `get it', in \citet{To:2020}'s words, and have the relevant knowledge and experience for navigating specific challenges in the discriminatory incident. The social interactions should then be guided around student needs in the context of the incident, \eg by building up on \citet{OLeary:2018}'s work. 
With respect to active coping, systems can allow specification of coping goals situated in the context of the incident, and enable tracking of coping behaviors is support of those goals. They can also recommend more advanced and diverse coping over time, \eg by appropriating \citet{Schroeder:2020}'s Goal Directed framework. 

\subsection{Proactive Interventions: Reducing Risks; Expanding Resources}
\label{sec:discussion-guide-proactive}
Interventions that strengthen students' access to resources and improve culture to reduce exposure to \discrimination, particularly a distressing one, can empower students and reduce the cumulative burden of the events. They can do so while placing some of the responsibility for change on others' shoulders. Since these interventions represent opportunities for structural change, we focus primarily on an academic context and efforts best spearheaded by educational institutions in the suggestions below\rev{}{, which underscore the need for improving mentoring, content access, and venue / program design through technologies.}

Interventions should help to establish connections to successful peers and advisors who have faced challenges similar to those faced by the students (\eg taking one's first computer science class with little preparatory high school classwork). Due to shared experiences, these mentors better relate to student needs and can more effectively act as role models \cite{Thomas:2001}. 
Such social resources may help students avoid or effectively resolve perilous situations when such situations arise.
Technology has already been adopted for e-mentoring \cite{Guy:2002, Direnzo:2013, Trainer:2017} but can go beyond providing a medium, \eg by connecting mentees to appropriate mentors to enhance mentee engagement \cite{Steinmacher:2012}. 
Future research should expand this existing work by additionally accounting for shared backgrounds and experiences in matching mentors and mentees, and by exploring e-mentoring models and processes 
for different student needs. 

Interventions can also connect students to other resources, many of which are unevenly used \cite{Hoskins:2005}. 
Personalized content curation and delivery, \eg consistent with schedules, attention resources, and mood, is a promising venue for interventions in this direction. For example, students could get notified of tutoring sessions, particularly for the courses they are struggling in. These notifications could maximize engagement by being delivered according to student schedule (\eg on fragmented days) or state (\eg when students are cognitively prepared \cite{Pielot:2015}). 

Consistent with past literature \cite{Panter:2008}, more than half of the incidents discussed in interviews or reported in experience sampling happened in academic settings, either in direct communication with academic personnel (\ie instructor, TA, or advisor) or in university-designed venues (\ie classrooms, study sessions, or labs). 
Since computer-mediated communication is an important aspect of educational programs, there may be opportunities for systems to check in about discouraging and discriminatory language just as there are now automated checks that raise concerns when a possible accessibility concern is detected \cite{AccessibilityScanner:Apple, AccessibilityScanner:Google,AccessibilityScanner:Blackboard,pendergast2015leveraging}. Moreover, systems can facilitate diverse participation, \eg through raising awareness of class dynamics and providing personalized recommendations by building up on such work as \cite{Samrose:2021} in a classroom context.

%% file: 8-conclusion.tex
\section{Conclusions and future work}
\label{sec:conclusion}
The purpose of this research was to identify directions and requirements for technologies that support students who face discrimination. To this end, we characterized \discrimination\ experiences by studying how they unfold. We identified factors that influence the experience: perils, beliefs, and coping resources. 
We examined the findings through the extended stress processing lens and \rev{}{offered a disciplined approach to identifying} design opportunities and requirements. Specifically, we \rev{}{highlighted that} addressing \discrimination\ should go beyond in-the-moment emotional comfort 
and conceptualized incident-specific and proactive design paradigms \rev{}{that improve other key elements of the experience}. We discussed a range of opportunities and requirements for mental health, social, and educational systems within these paradigms and argued they can facilitate meaningful support. 

A feature, but also a limitation, of our approach is the breadth of \discrimination\ incidents we included in our data. Successful intervention design will likely require similar work to take place with specific \rev{}{social} groups to understand identity-specific needs. 

\label{sec:discussion-future}

Overall, we hope our work inspire research in multiple areas including mental health, social systems, and education, with a specific focus on \rev{}{supporting} individuals against the harm of \discrimination. 

%% file: 9-appendix.tex
\section*{Appendix}
\label{sec:appendix}

\begin{table*}[h]
\centering
\resizebox{0.9\textwidth}{!}{%
\begin{tabular}{|lll|}
\hline
praying or meditating & keeping it sealed up & letting it out \\
crying & mindless scrolling & lashing out at people \\
walking & turning to work & avoiding people and places \\
exercising & sleeping & ignoring the situation \\
breathing exercises, yoga & binge watching & searching for less negative explanations \\
planning & playing games & looking at the bright side \\
taking steps to change the situation & shopping & finding humor in the situation \\
seeking help or advice & eating & acknowledging the situation \\
challenging the person & drinking & ruminating \\
advocating for change & smoking, using drugs & self-blaming \\
seeking emotional support & giving up &  \\ \hline
\end{tabular}%
}
\caption[Coping Strategies]{List of Strategies Presented on Coping Cards (\eg \fig{fig:coping-cards}).}
\label{tab:coping-cards}
\vspace{-0.35in}
\end{table*}